\documentclass[fleqn,12pt]{article}
\usepackage{graphics,epsfig,float,mathrsfs,color}
\usepackage{appendix}

\newcommand{\be}[1]{\begin{equation} \label{(#1)}}
\newcommand{\ee}{\end{equation}}
\newcommand{\baq}[1]{\begin{eqnarray} \label{(#1)}}
\newcommand{\eaq}{\end{eqnarray}}

\newcommand{\rf}[1]{(\ref{(#1)})}
\newcommand{\BR}{\textrm{BR}}

\def\lsim{\raise0.3ex\hbox{$\;<$\kern-0.75em\raise-1.1ex\hbox{$\sim\;$}}}
\def\gsim{\raise0.3ex\hbox{$\;>$\kern-0.75em\raise-1.1ex\hbox{$\sim\;$}}}

\newcommand{\AddrAHEP}{
\it AHEP Group, Institut de F\'isica Corpuscular - C.S.I.C. \\
Universitat de Val\`encia, Edifici Instituts d'Investigaci\'o \\ 
Apt. 22085, E-46071 Val\`encia, Spain \\}

\newcommand{\AddrCFTP}{
\it Departamento de F\'isica and CFTP, Instituto Superior T\'ecnico, \\
Avenida Rovisco Pais 1, 1049-001 Lisboa, Portugal \\}

\begin{document}

\begin{flushright}
 IFIC/09-48\\
 CFTP/09-033
\end{flushright}

\begin{center}  
  \textbf{\large 
Minimal Supersymmetric Inverse Seesaw: Neutrino masses, lepton 
flavour violation and LHC phenomenology}\\[10mm]

{M.~Hirsch${}^1$, T.~Kernreiter${}^2$, J.~C.~Rom\~ao${}^2$, 
and A.~Villanova del Moral${}^2$
} 
\vspace{0.3cm}\\ 

$^1$\AddrAHEP
$^2$ \AddrCFTP

\end{center}

\vskip2cm
\noindent

\begin{abstract}
\noindent
We study neutrino masses in the framework of the supersymmetric inverse 
seesaw model. Different from the non-supersymmetric version a minimal 
realization with just one pair of singlets is sufficient to explain 
all neutrino data. We compute the neutrino mass matrix up to 1-loop 
order and show how neutrino data can be described in terms of the model 
parameters. We then calculate rates for lepton flavour violating (LFV) 
processes, such as $\mu \to e \gamma$, and chargino decays to singlet 
scalar neutrinos. The latter decays are potentially observable at the LHC 
and show a characteristic decay pattern dictated by the same parameters 
which generate the observed large neutrino angles.

\end{abstract}

\newpage 

%-----------------------------------------------------------------------------
\section{Introduction}
%-----------------------------------------------------------------------------

Currently there are only very few indications for physics beyond the 
standard model (SM), the most important ones coming from neutrino physics 
and cosmology. On the one hand, neutrino oscillation experiments 
\cite{Fukuda:1998mi} have shown that at least two neutrinos have 
non-zero masses and that mixing angles in the lepton sector are 
surprisingly large \cite{Maltoni:2004ei}. On the other hand, data 
from the WMAP satellite \cite{Spergel:2003cb,Komatsu:2008hk} and 
large scale structure formation \cite{Tegmark:2006az} have provided 
convincing evidence for the existence of non-baryonic dark matter. 

In this paper we study a minimal supersymmetric version of the 
inverse seesaw \cite{Mohapatra:1986bd}. The model is capable to 
explain all neutrino data with only one pair of singlet superfields. 
It contains a new dark matter candidate - the scalar singlet - 
which can give the correct relic density \cite{Arina:2008bb} and 
it gives potentially testable predictions for both, supersymmetric 
phenomenology at the LHC and low energy lepton flavour violating 
decays, such as $\mu\to e \gamma$.

Neutrino masses are not part of the SM, but models which can explain 
oscillation data are quite easily constructed. Indeed, it was pointed 
out already in \cite{Weinberg:1979sa} that for Majorana neutrinos 
the mass matrix is described by a unique dimension-5 operator,
\begin{equation}\label{eq:dim5}
m_{\nu} = \frac{f}{\Lambda} (H L) (H L).
\end{equation}
All models which reduce to the SM particle content at low energy are 
merely different realizations of this operator and at tree-level there 
are just three basic contractions which give rise to eq.~(\ref{eq:dim5}) 
\cite{Ma:1998dn}. The literature is completely dominated by only one of 
them, based on the exchange of heavy singlets~\cite{Minkowski:1977sc,seesaw}. 
This is now commonly called the (type-I) seesaw mechanism. 

In type-I seesaw the smallness of the observed neutrino masses is 
attributed to the large mass of the singlets ($\nu^c$) and for 
$f \sim {\cal O}(1)$ current neutrino data indicates $\Lambda \simeq 
10^{15}$ GeV. Obviously, if this ansatz is the correct explanation for 
neutrino masses, it will never be directly tested\footnote{Dirac 
neutrinos can just as easily explain oscillation data. However, Dirac 
neutrinos require Yukawa couplings of order ${\cal O}(10^{-12})$ or 
smaller, thus there is no conceivable experimental phenomenology outside 
the neutrino sector for Dirac neutrinos either.}. However, the smallness 
of $m_{\nu}$ could be understood as well, if $f$ is $f \ll 1$. The 
classical examples for this situation are radiative neutrino mass 
models \cite{Cheng:1980qt,Zee:1980ai,Babu:1988ki}. 

In the inverse seesaw model \cite{Mohapatra:1986bd} the particle content 
of the SM is extended by one or more pairs of singlets, call them $\nu^c$ 
and $S$, which form ``heavy'' pseudo-Dirac pairs. The smallness of 
$m_{\nu}$ is then attributed to a small lepton number breaking parameter, 
$\mu_S$. The smallness of this parameter is natural in the t'Hooft 
sense \cite{'tHooft:1979bh}, since a vanishing $\mu_S$ restores 
a symmetry of the theory. Similar to the ordinary type-I seesaw, in 
the inverse seesaw only one non-zero neutrino mass for the light 
neutrino fields is generated for each pair of singlets. A non-supersymmetric 
inverse seesaw thus needs at least two pairs of singlets to explain 
neutrino oscillation data  \cite{Malinsky:2009df}. As we show below, 
in a {\em supersymmetric} inverse seesaw one pair of singlets is sufficient 
to explain the experimental data. In such a minimal supersymmetric 
inverse seesaw model (MSISM) one neutrino mass is generated at tree-level, 
while a second non-zero mass is due to the scalar neutrino-antisneutrino 
loop \cite{Hirsch:1997vz}. The scheme we consider is reminiscent 
of bilinear R-parity violation, which is also of the hybrid ``tree + loop'' 
type \cite{Hirsch:2000ef}.

Supersymmetrizing the inverse seesaw offers additional advantages\footnote{A 
supersymmetric extension of the SM which adds only 
singlets inherits all the standard arguments in favour of SUSY, 
such as providing a (technical) solution to the gauge hierarchy 
problem, gauge coupling unification, etc.}. 
Cosmology requires the existence of a non-baryonic dark matter (DM) 
candidate and SUSY with conserved R-parity offers a WIMP candidate 
in the form of the lightest supersymmetric particle (LSP), for 
reviews see for example \cite{Jungman:1995df,Bertone:2004pz}. 
In the minimal supersymmetric extension of the standard model (MSSM) 
only the lightest neutralino remains as a CDM candidate, since left 
sneutrinos have been ruled out as cold dark matter by a combination 
of experimental data from LEP and direct detection experiments 
\cite{Falk:1994es}. Right sneutrinos could be the CDM, however, for 
in the case of pure Dirac neutrinos as well as in the case of the standard type-I seesaw 
Majorana neutrinos, the sneutrinos are expected to have such small 
couplings to all ordinary particles that they can not be {\em thermally} 
produced dark matter. Non-thermal right sneutrino DM has been discussed 
in \cite{Asaka:2005cn,Gopalakrishna:2006kr}. Right sneutrinos could be 
thermalized in the early universe, if they have (a) enlarged left-right 
mixing \cite{ArkaniHamed:2000bq,Arina:2007tm}; (b) a large quartic coupling 
to the Higgs fields \cite{Deppisch:2008bp}; (c) an extra 
$U'(1)$ under which sneutrinos are charged \cite{Lee:2007mt} or (d) 
within the NMSSM, if the sneutrinos have a large coupling to the 
NMSSM singlet \cite{Cerdeno:2009dv}. In the supersymmetric inverse seesaw, 
which we consider here, the singlet scalars are expected to be 
thermal cold dark matter candidates \cite{Arina:2008bb}, since the 
neutrino Yukawa couplings are much larger than in the standard type-I 
seesaw. 

The large Dirac neutrino couplings lead necessarily also to non-zero 
lepton flavour violating processes, such as $\mu\to e \gamma$ and 
LFV supersymmetric particle decays. We therefore calculate BR($\ell_j \to 
\ell_i + \gamma$) and compare the expected rates with experimental 
sensitivities. If SUSY particles are light enough to be produced at 
the LHC, the new singlet states can appear in the decay chains, 
potentially altering the phenomenology. This is especially important 
in case one of the singlets is the LSP. We therefore also calculate 
the decays $\chi^+_1 \to \ell_i + {\tilde N}_a$, where ${\tilde N}_a$ 
stands for a scalar neutrino. The flavour of the lepton in these 
decays can be tagged and traces the lepton flavour violating couplings 
of the sneutrinos. We show how these LFV couplings are related to the 
observed neutrino angles in the theoretically preferred part of the 
parameter space.

The rest of this paper is organized as follows. In the next section 
we outline the model and calculate the neutrino masses at 1-loop order. 
Section~\ref{sec:fit} then presents some approximate formulas 
for neutrino masses and mixing angles, which allow to understand how 
the model can explain the experimental data. We then turn to 
phenomenology in section~\ref{sec:pheno}. We calculate the decays of 
the lightest chargino to leptons plus scalar neutrino, assuming the 
(singlet) sneutrinos are the LSP. We compare the expected signals with 
limits on parameters imposed by BR($\mu \to e + \gamma$). We then close 
with a short summary. Some formulas for the calculations of loops and 
LFV decays are relegated to the appendix.

%-----------------------------------------------------------------------------
\section{Minimal supersymmetric inverse seesaw} 
%-----------------------------------------------------------------------------
\label{sec:theory}

\subsection{The model}
\label{subsec:SPot}

The model is defined by the superpotential of the MSSM extended 
by a pair of singlet fields, $\widehat{\nu}^c$ and $\widehat{S}$ 
with lepton numbers assigned to be $-1$ and 1, respectively. 
The total superpotential contains then three additional terms 
\cite{Arina:2008bb}
\be{eq:SuperPot}
{\mathcal W}={\mathcal W}_{\rm MSSM}+\varepsilon_{ab}h^i_\nu\widehat{L}^a_i
\widehat{\nu}^c\widehat{H}_u^b+M_R\widehat{\nu}^c\widehat{S}+
\frac{1}{2}\mu_S\widehat{S}\widehat{S}~.
\ee
Note that, in the limit where $\mu_S\to 0$, lepton number is conserved 
and that the parameter $M_R$ does not violate lepton number. We introduce 
only one generation of $\widehat{\nu}^c$ and $\widehat{S}$.  This model 
is thus the minimal supersymmetric inverse seesaw model (MSISM) capable 
of explaining neutrino data. Previous works used three generations of 
singlets, see e.g. \cite{Arina:2008bb,Deppisch:2004fa}. The model conserves 
$R-$parity, and as a consequence, the lightest SUSY particle is stable.

With the additional singlet fields the soft SUSY breaking Lagrangian 
is specified by
\baq{eq:softSUSY}
-{\mathcal L}_{\rm soft} &=& -{\mathcal L}^{\rm MSSM}_{\rm soft} 
         +  m^2_{\nu^c} \widetilde\nu^{c\dagger} \widetilde\nu^c
         +m^2_S \widetilde S^{\dagger} \widetilde S\nonumber\\[2mm]
    {}&& + \left(\varepsilon_{ab} A^i_{h_\nu}
                \widetilde L^a_i \widetilde\nu^c H_u^b +
                B_{M_R} \widetilde\nu^c \widetilde S 
      +\frac{1}{2}B_{\mu_S}\widetilde S \widetilde S
      +{\rm h.c.}\right)~,
\eaq
where ${\mathcal L}^{\rm MSSM}_{\rm soft}$ contains the usual soft 
SUSY breaking terms of the MSSM. The parameter $B_{\mu_S}$ is the 
analogue of the lepton number violating parameter $\mu_S$ in the 
superpotential. The model thus includes two parameters which violate 
lepton number, both will necessarily contribute to the (Majorana) 
neutrino mass matrix.

%-----------------------------------------------------------------------------
\subsection{Tree--level neutrino and sneutrino masses}
%-----------------------------------------------------------------------------

From eq.~\rf{eq:SuperPot} we obtain the mass matrix of the neutral 
fermion fields, which, in the basis ($\nu_e,\nu_\mu,\nu_\tau,\nu^c,S$), 
reads  
\be{eq:Neutrino}
M^\nu=\left(
\begin{array}{ccccc}
0 & 0 & 0 & m_{D_1}& 0 \\[2mm]
0 & 0 & 0 & m_{D_2}& 0 \\[2mm]
0 & 0 & 0 & m_{D_3}& 0 \\[2mm]
m_{D_1}& m_{D_2}& m_{D_3} & 0 & M_R\\[2mm]
0 & 0 & 0 & M_R & \mu_S 
\end{array} \right),
\ee
where $m_{D_i}\equiv h_\nu^i v_u$ ($i=1,2,3$), with $v_u$ being the 
vacuum expectation value of the Higgs field, $\langle H^0_u\rangle$. 
For $m_{D_i}\ll M_R$, one obtains the effective ($3\times 3$) mass matrix 
of the light neutrinos in the seesaw approximation:
\be{eq:SeeSaw}
(M^\nu_{\rm mass})_{ij}= \frac{\mu_S}{M_R^2}~m_{D_i}m_{D_j}~.
\ee
The lepton number violating parameter $\mu_S$ controls the absolute scale 
of the neutrino masses. Eq.~\rf{eq:SeeSaw} shows manifestly the projective 
nature of the light neutrino mass matrix. Thus only one neutrino mass 
is non-zero at tree-level. However, this result is true in general and 
does not depend on the seesaw approximation. Note also, that if $m_{D_i}$ 
is of the same order as $M_R$ the correct eigenvalue is found by 
replacing $M_R^2 \to M_R^2 + \sum m_{D_i}^2$ in eq.~\rf{eq:SeeSaw}.

The neutrino mass matrix in eq.~\rf{eq:SeeSaw} is diagonalized by 
an unitary transformation in the standard way 
\be{eq:diagNeu}
U^{{\rm tr} T}~M^\nu_{\rm mass}~U^{\rm tr}= {\rm diag}(0,0,m_{\nu_3})~.
\ee
In order to obtain a second non-vanishing neutrino mass eigenvalue,
loop corrections must be included. In this context it is amusing to 
note that in the non-SUSY case the inverse seesaw requires two copies 
of the singlet fields, $\nu^c_i,S_i$ ($i=1,2$), in order to give rise to 
a viable neutrino mass matrix \cite{Malinsky:2009df}, even after loop 
corrections are taken into account.

Assuming CP conservation the 10$\times$10 sneutrino mass matrix 
can be decomposed into two 5$\times$5 matrices for the
CP-even, $\phi^R=(\widetilde\nu^R_e,\widetilde\nu^R_\mu,\widetilde\nu^R_\tau,
\widetilde\nu^{cR},\widetilde S^R)$, and CP-odd scalar fields,
$\phi^I=(\widetilde\nu^I_e,\widetilde\nu^I_\mu,\widetilde\nu^I_\tau,$ $
\widetilde\nu^{cI},\widetilde S^I)$, respectively, and reads
\be{eq:Sneutrino}
{\mathcal L}_{\tilde\nu}=\frac{1}{2}~(\phi^R,\phi^I) \left(
\begin{array}{cc}
M^2_+ &  0 \\
0 &  M^2_- \\
\end{array} \right)\left(
\begin{array}{c}
\phi^R \\
\phi^I 
\end{array}\right).
\ee
The two mass matrices $M^2_\pm$ are given by \cite{Arina:2008bb}
\be{eq:SneuDecom}
\!\!\!\!\!\!\!\!\!\!\!\!\!\!\!\!\left(\begin{array}{ccc}
(M^2_{\tilde L_i}+\frac{1}{2}m_Z^2 \cos2\beta+m^2_{D_i})\delta_{ij} & 
\pm(A^j_{h_\nu}v_u-\mu~ m_{D_j} \cot\beta) & m_{D_j} M_R\\[2mm]
\pm(A^i_{h_\nu}v_u-\mu~ m_{D_i} \cot\beta) & 
 m^2_{\nu^c}+M_R^2+\sum^3_{k=1}m^2_{D_k} &
 \mu_S M_R\pm B_{M_R}\\[2mm]
m_{D_i} M_R & \mu_S M_R\pm B_{M_R} &m_S^2+\mu^2_S+M_R^2\pm B_{\mu_S}
\end{array} \right),
\ee
where we use a compact form to write these matrices with the index
$i$ for the row and the index $j$ for the column, $i,j=1,2,3$. 
The real symmetric mass matrix in eq.~\rf{eq:Sneutrino} can be
diagonalized by a 10$\times$10 orthogonal matrix as follows
\be{eq:Sneumix2}
G ~ M^2_{\tilde\nu}~ G^T=
{\rm diag}(m^2_{\tilde N_1},\dots, m^2_{\tilde N_{10}})~,
\ee
with $m^2_{\tilde N_1}<\dots< m^2_{\tilde N_{10}}$.
Diagonalizing the mass matrices for the CP-even and CP-odd
mass matrices $M^2_\pm$ separately by 
\be{eq:Sneumix}
G_\pm~ M^2_\pm~ G^T_\pm={\rm diag}(m^2_{\tilde\nu^\pm_1},
         \dots, m^2_{\tilde\nu^\pm_5})~,
\ee
leads to a parametrization which is useful for a qualitative discussion 
of the parameter dependence of the neutrino mass matrix 
which we wish to address below. 

%
%-----------------------------------------------------------------------------
\subsection{Neutrino mass matrix at 1-loop order}
%-----------------------------------------------------------------------------

We now compute the 1-loop radiative corrections to the neutrino mass matrix.
The amplitude for the loop contributions to the neutrino self-energy can be
generically written as\footnote{In order to make our results more easily 
comparable with the case of the standard supersymmetric type-I seesaw, 
we closely follow the notation of \cite{Dedes:2007ef}.}
\be{eq:Ampl}
-i\Sigma_{\nu_m\nu_n}(p)=-i
\left[(p\!\!\!/~\Sigma^{mn}_V(p^2)+\Sigma^{mn}_S(p^2)) P_L+
(p\!\!\!/~\Sigma^{mn*}_V(p^2)+\Sigma^{nm*}_S(p^2)) P_R
\right].
\ee
Clearly, the self-energy functions $\Sigma^{mn}_{S,V}(p^2)$ must
be symmetric with interchanging their indices due to the Majorana 
nature of the neutrinos. The 1-loop corrected neutrino mass matrix 
is given by
\be{eq:LoopNeu}
M^{{\rm 1-loop}}_{mn}= 
m_{\nu_m}(Q)\delta_{mn}+{\rm Re}\left[\Sigma^{mn}_S(p^2)
+m_{\nu_m}\Sigma^{mn}_V (p^2)\right]_{\Delta=0},
\ee
where $m_{\nu_m}=(0,0,m_{\nu_3})$ and the self-energy functions 
$\Sigma^{mn}_{S,V}(p^2)$ are evaluated at $p^2=m^2_{\nu_3}$, which 
is tiny compared to the masses of the particles in the loop, and
in excellent approximation can be set to zero. The 
divergences in eq.~\rf{eq:LoopNeu} are removed, using the minimal 
subtraction scheme, i.e. by setting the parameter 
$\Delta\equiv 2/(4-d)-\gamma_E+\log 4\pi=0$. Here, as usual, $d$ 
is the number of space-time dimensions, $\gamma_E$ is the Euler 
constant, and $Q$ is the renormalization scale at which the input 
parameters are defined.

The 1-loop improved neutrino mass matrix in eq.~\rf{eq:LoopNeu} is 
then diagonalized by an unitary matrix denoted as $U^{\rm 1-loop}$. 
The neutrino mixing matrix relating the flavor basis ($\nu_\alpha$)  
and the mass eigenbasis ($\nu_i$) of the light neutrinos is then 
given by
\be{eq:NeuMixing}
\nu_\alpha = (U^{\rm tr} U^{\rm 1-loop})_{\alpha i}~\nu_i 
         \equiv U^{\nu}_{\alpha i}~\nu_i~.
\ee

There are two different types of 1-loop diagrams. One class of 
diagrams exchanges Higgses and neutrinos. As we show in detail 
in the appendix, the flavour structure of this loop repeats the 
flavour structure of the tree-level mass matrix, eq. \rf{eq:SeeSaw}, 
and thus only renormalizes $m_{\nu_3}$. More important is the scalar 
neutrino-antisneutrino loop, since it implements a new flavor structure 
(besides $h^i_\nu$) and thus generates a second non-zero neutrino mass. 
The new flavor structure is due to the trilinear couplings $A^i_{h_\nu}$, 
see eq.~\rf{eq:softSUSY}.

The relevant interaction for the calculation of the self-energy functions 
is the sneutrino-neutralino-neutrino interaction, which is given by the 
Lagrangian
\be{eq:Lagrange}
{\mathcal L}_{\nu\chi^0\tilde\nu}=
\bar{\widetilde\chi}_j^0(A^R_{mjb}P_R+A^L_{mjb} P_L)\nu_m \widetilde N_b
+{\rm h.c.}~,
\ee
with
\begin{eqnarray}
A^R_{mjb} &=&-\frac{1}{\sqrt{2}}h_\nu^i U^{\rm tr}_{im} N_{j4}(G_{b4}-i G_{b9})
~,
\label{eq:couplR} \\[2mm]
A^L_{mjb} &=&-\frac{g}{2}(N^*_{j2}-\tan\theta_W N^*_{j1})
(G_{bi}-i G_{b(i+5)})U^{\rm tr}_{im}~,
\label{eq:couplL}
\end{eqnarray}
where $g$ is the $SU(2)_L$ gauge coupling and $\theta_W$ is the
weak mixing angle, respectively, and $N$ is the unitary 4$\times$4 
neutralino mixing matrix, which diagonalizes the neutralino mass 
matrix by $N^* M_{\chi^0} N^{-1}={\rm diag}(m_{\chi^0_1},\dots,m_{\chi^0_4})$, 
with $m_{\chi^0_j}>0$. The sneutrino mass matrix and diagonalization 
have been discussed in the previous section. 

The calculation of the self-energy functions then yields
\begin{eqnarray}
\!\!\!\!\!\!\!\!\!\Sigma^{mn}_{S2} &=& \frac{-m_{\chi^0_j}}{(4\pi)^2}
\left[
A^L_{mjb}A^L_{njb}+A^{R*}_{mjb}A^{R*}_{njb}+A^{R*}_{mjb}A^L_{njb}
+A^L_{mjb}A^{R*}_{njb}
\right] B_0(m^2_{\chi^0_j},m^2_{\tilde N_b})\label{eq:AmplS2},\\[2mm]
\!\!\!\!\!\!\!\!\!\Sigma^{mn}_{V2} &=& \frac{-1}{(4\pi)^2}
\left[
A^{L*}_{mjb}A^L_{njb}+A^R_{mjb}A^{R*}_{njb}+A^{L*}_{mjb}A^{R*}_{njb}
+A^{R}_{mjb}A^L_{njb}
\right] B_1(m^2_{\chi^0_j},m^2_{\tilde N_b}).
\label{eq:AmplV2}
\end{eqnarray}
In the limit that the right-chiral couplings $A^R_{mjb}$, 
eq.~(\ref{eq:couplR}), are omitted, the result of the 
sneutrino-neutralino loop calculation \cite{Dedes:2007ef}
in the standard type-I (SUSY) seesaw is recovered. In the 
type-I seesaw the couplings $A^R_{mjb}$ are negligible 
because of the tiny mixing among left-handed and right-handed 
sneutrino states, which are separated by a large mass hierarchy.
On the other hand, in the inverse seesaw left-handed and right-handed 
sneutrino states have similar mass scales, such that the contribution 
of couplings $A^R_{mjb}$ is relevant. As has been noted \cite{Dedes:2007ef}, 
the functions $\Sigma^{mn}_{S2}$ are UV finite, as the $\Delta$ part in
the loop function $B_0(m^2_{\chi_j},m^2_{\tilde N_b})$ drops out,
because of the orthogonality of $G$. For the same reason $\Sigma^{mn}_{S2}$ 
is also independent of the renormalization scale $Q$.
The off-diagonal functions $\Sigma^{m\neq n}_{V2}$ are finite, as expected, 
and only the 33-element of the diagonal elements 
$\Sigma^{mm}_V$ gives a non-vanishing contribution to the neutrino mass
matrix, because of the second
term in eq.~\rf{eq:LoopNeu}, where $\Sigma^{33}_{V2}$ retains a dependence
on $\Delta$ and $Q$.
Numerically we find that the contributions due to $\Sigma^{mn}_{V2}$, which 
are multiplied by the small neutrino mass, see eq.~\rf{eq:LoopNeu}, 
are much smaller than the contributions due to $\Sigma^{mn}_{S2}$
and can be safely neglected in general.
The self-energy functions in eq.~\rf{eq:LoopNeu} is then given by the sum 
of the neutrino-Higgs (see appendix) and sneutrino-neutralino contributions 
\be{eq:}
\Sigma^{mn}_{S}(p^2)=\Sigma^{mn}_{S1}(p^2)+\Sigma^{mn}_{S2}(p^2)~,\quad
\Sigma^{mn}_{V}(p^2)=\Sigma^{mn}_{V1}(p^2)+\Sigma^{mn}_{V2}(p^2)~.
\ee
%

%---------------------------------------------------------------------------
\section{Approximate expressions for neutrino masses and fit to 
experimental data}
%---------------------------------------------------------------------------
\label{sec:fit}

The lepton number violating parameters $\mu_S$ and $B_{\mu_S}$ govern the 
scale of neutrino physics.  $B_{\mu_S}$ essentially controls the size 
of the loop contributions, while $\mu_S$ is restricted due to the 
tree-level neutrino mass (and thus plays only a sub-leading role in 
the loops). However, only in the limit where both lepton number violating 
parameters vanish, i.e. $\mu_S,B_{\mu_S}\to 0$, the masses of the CP-even 
and CP-odd scalars are pairwise equal, i.e. 
$m^2_{\tilde\nu^+_1}=m^2_{\tilde\nu^-_1},\dots,m^2_{\tilde\nu^+_5}
=m^2_{\tilde\nu^-_5}$. In this limit there is then a complete cancellation 
between the contributions of the CP-even and CP-odd scalar loops 
\cite{Hirsch:1997vz}. 

The role of other model parameters can be understood with the help of 
the following approximate relations. In the flavour basis one can 
write the 1-loop contribution to the neutrino mass matrix as 
\be{eq:GenericNeu}
M_{\nu}^{\rm 1-loop} = a~\varepsilon_m\varepsilon_n+b~(\varepsilon_m \delta_n+\delta_m\varepsilon_n)+
c~\delta_m\delta_n~,
\ee
with the vectors $\varepsilon_m$ and $\delta_m$ defined as 
\baq{eq:SneuApp}
\varepsilon_m &\equiv &  m_{D_m} M_R~,\\[2mm]
\delta_m &\equiv & A^m_{h_\nu}v_u-\mu~ m_{D_m}\cot\beta~~,
\eaq
with $a,b$ and $c$ being coefficients that depend on all other model 
parameters, see below. It is important to note that the $\epsilon_m$ 
have the same flavour dependence as the tree-level neutrino mass 
contribution, thus the vectors $\varepsilon_m$ and $\delta_m$ must 
not be aligned in order to explain neutrino data correctly. Note also 
that the structure in \rf{eq:GenericNeu} is only strictly true, if the 
soft SUSY breaking parameters $M_{\tilde L_i}$ are equal for all 
generations, i.e. $M_{\tilde L_1}=M_{\tilde L_2}=M_{\tilde L_3}$. 
Otherwise the new flavour structure introduced by $M_{\tilde L_i}$ should be taken into account. 

The coefficients $a,b$ and $c$ are found with the help of an 
approximative diagonalization of the scalar neutrino mass matrices. 
We give below the formulas for the case $M_{\tilde L_i}\gg M_R$ 
(where $M_R$ for simplicity stands for all parameters of the singlet 
sector) since this case is phenomenologically more interesting, as 
explained in the next section. Formulas for $M_{\tilde L_i}\ll M_R$ 
can be found easily. Expanding $A^{L/R}_{mjb}$ in the ``small'' 
parameters $\varepsilon_m$ and $\delta_m$ the coefficients read 
\baq{eq:Coeffa}
a & = &-\sum_j \frac{m_{\chi^0_j}}{(4\pi)^2}\left[
\frac{a_L^{j2}}{\widehat M^4_{\tilde L}}
\left(\cos^2\theta_+ B_0(m^2_{\chi^0_j},m^2_{\tilde\nu^+_2})-
\cos^2\theta_- B_0(m^2_{\chi^0_j},m^2_{\tilde\nu^-_2}) \right.\right.
\nonumber\\[2mm]
{}&& \left.\left.
\hskip+2.9cm +\sin^2\theta_+ B_0(m^2_{\chi^0_j},m^2_{\tilde\nu^+_1})-
\sin^2\theta_- B_0(m^2_{\chi^0_j},m^2_{\tilde\nu^-_1})\right)\right.
\nonumber\\[2mm]
{}&& \left. 
\hskip+1.5cm
+\frac{a_R^{j2*}}{M^2_R v_u^2}\left(\cos^2\theta_+ 
B_0(m^2_{\chi^0_j},m^2_{\tilde\nu^+_1})-
\cos^2\theta_- B_0(m^2_{\chi^0_j},m^2_{\tilde\nu^-_1})
\right.\right.\nonumber\\[2mm]
{}&& \left.\left.
\hskip+2.9cm +\sin^2\theta_+ B_0(m^2_{\chi^0_j},m^2_{\tilde\nu^+_2})-
\sin^2\theta_- B_0(m^2_{\chi^0_j},m^2_{\tilde\nu^-_2})\right)\right.
\nonumber\\[2mm]
{}&& \left.
\hskip+0.9cm
+\frac{a_R^{j*}a_L^j}{\widehat M^2_{\tilde L} M_R v_u}
\left(
\sin2\theta_- (B_0(m^2_{\chi^0_j},m^2_{\tilde\nu^-_2})-
B_0(m^2_{\chi^0_j},m^2_{\tilde\nu^-_1}))\right.\right.\nonumber\\[2mm]
{}&& \left.\left.
\hskip+2.9cm
+\sin2\theta_+ (B_0(m^2_{\chi^0_j},m^2_{\tilde\nu^+_2})-
B_0(m^2_{\chi^0_j},m^2_{\tilde\nu^+_1}))
\right)\right]~,
\eaq
\baq{eq:Coeffb}
b &=&  \sum_j \frac{m_{\chi^0_j}}{(4\pi)^2}
\left[\frac{a_R^{j*}a_L^j}{\widehat M^2_{\tilde L} M_R v_u}
\left(\cos^2\theta_+ B_0(m^2_{\chi^0_j},m^2_{\tilde\nu^+_1})-
\cos^2\theta_- B_0(m^2_{\chi^0_j},m^2_{\tilde\nu^-_1}) \right.\right.
\nonumber\\[2mm]
{}&& \left.
\hskip+3.6cm +\sin^2\theta_+ B_0(m^2_{\chi^0_j},m^2_{\tilde\nu^+_2})-
\sin^2\theta_- B_0(m^2_{\chi^0_j},m^2_{\tilde\nu^-_2})\right)\nonumber\\[2mm]
{}&& \left.
\hskip+2.6cm 
+\frac{a_L^{j2}}{\widehat M^4_{\tilde L}}
\left(
\sin2\theta_- (B_0(m^2_{\chi^0_j},m^2_{\tilde\nu^-_2})-
B_0(m^2_{\chi^0_j},m^2_{\tilde\nu^-_1}))\right.\right.\nonumber\\[2mm]
{}&& \left.\left.
\hskip+3.6cm
+\sin2\theta_+ (B_0(m^2_{\chi^0_j},m^2_{\tilde\nu^+_2})-
B_0(m^2_{\chi^0_j},m^2_{\tilde\nu^+_1}))
\right)\right]~,
\eaq
\baq{eq:Coeffc}
c &=&  -\sum_j \frac{m_{\chi^0_j}}{(4\pi)^2}\left[
\frac{a_L^{j2}}{\widehat M^4_{\tilde L}}
\left(\cos^2\theta_+ B_0(m^2_{\chi^0_j},m^2_{\tilde\nu^+_1})-
\cos^2\theta_- B_0(m^2_{\chi^0_j},m^2_{\tilde\nu^-_1}) \right.\right.
\nonumber\\[2mm]
{}&& \left.\left.
\hskip+2.9cm +\sin^2\theta_+ B_0(m^2_{\chi^0_j},m^2_{\tilde\nu^+_2})-
\sin^2\theta_- B_0(m^2_{\chi^0_j},m^2_{\tilde\nu^-_2})\right)\right]~,
\eaq
Here we have used the abbreviations 
$a_L^j=-g/2(N^*_{j2}-\tan\theta_W N^*_{j1})$,
$g_R^j=-1/\sqrt{2} N_{j4}$ and $\widehat M^2_{\tilde L}=M^2_{\tilde L}+
\frac{1}{2}m_Z^2\cos2\beta$.
The mixing angles $\theta_{\pm}$ diagonalize the 2$\times$2 sub-matrices
of the $(\widetilde\nu^{cR},\widetilde S^R)$ and 
$(\widetilde\nu^{cI},\widetilde S^I)$ systems, respectively, and are given by 
\baq{eq:thetap}
\hskip-1cm
\cos\theta_+ =
\frac{-B_{M_R}}{\sqrt{B_{M_R}^2+(m^2_{\tilde\nu^+_1}-m_{\nu^c}^2-M_R^2)^2}},~~
\sin\theta_+ =
\frac{m_{\nu^c}^2+M_R^2-m^2_{\tilde\nu^+_1}}
{\sqrt{B_{M_R}^2+(m^2_{\tilde\nu^+_1}-m_{\nu^c}^2-M_R^2)^2}},
\eaq
\baq{eq:thetam}
\hskip-1cm
\cos\theta_- =
\frac{B_{M_R}}{\sqrt{B_{M_R}^2+(m^2_{\tilde\nu^-_1}-m_{\nu^c}^2-M_R^2)^2}},~~
\sin\theta_- =
\frac{m_{\nu^c}^2+M_R^2-m^2_{\tilde\nu^-_1}}
{\sqrt{B_{M_R}^2+(m^2_{\tilde\nu^-_1}-m_{\nu^c}^2-M_R^2)^2}},
\eaq
where we have neglected the tiny $\mu_S$ term in the (2,2) entry and 
the $\sum_k m_{D_k}^2$ term (which is of higher order in the seesaw 
expansion) in the (1,1) entry. The corresponding mass eigenvalues are 
denoted by $m^2_{\tilde\nu^+_{1,2}}$ and $m^2_{\tilde\nu^-_{1,2}}$.
We stress that, in order to derive the analytic formulas for the sneutrino 
mixing angles $\theta_{\pm}$ we have implicitly assumed that the mixing 
of the singlet sneutrinos to the left sneutrinos is small, i.e. 
$\epsilon_m$ and $\delta_m$ are smaller than all other mass squared 
parameters of the problem. 

Let us consider the case where the mass eigenstates are close to the 
weak eigenstates, i.e. the mixing angles $\theta_{\pm}$ are close to 
0 or $\pi/2$. This corresponds to the parameter $B_{M_R}$ being small. 
It can be shown that the mass squared difference 
$m^2_{\tilde\nu^+_1}-m^2_{\tilde\nu^-_1}$ goes to zero for 
$\cos\theta_{\pm}\to 1$, while for $\cos\theta_{\pm}\to0$ the mass 
squared difference of the heavier states  approaches zero, i.e. 
$m^2_{\tilde\nu^+_2}-m^2_{\tilde\nu^-_2}\to 0$. Given this result, 
from eqs.~\rf{eq:Coeffa}-\rf{eq:Coeffc} one finds that in the limit 
$B_{M_R}\to 0$, only the coefficient $a$ is non-vanishing. 
For a viable neutrino mass matrix, however, we will need also 
a contribution from the last term in eq.~\rf{eq:GenericNeu} and 
this in turn requires therefore a sizeable $B_{M_R}$. 
 Note that this results holds true also for the reversed case, 
i.e. $M_{\tilde L_i}\ll M_R$.

\begin{figure}[htb]
  \centering
  \begin{tabular}{cc}
  \includegraphics[width=0.45\linewidth]{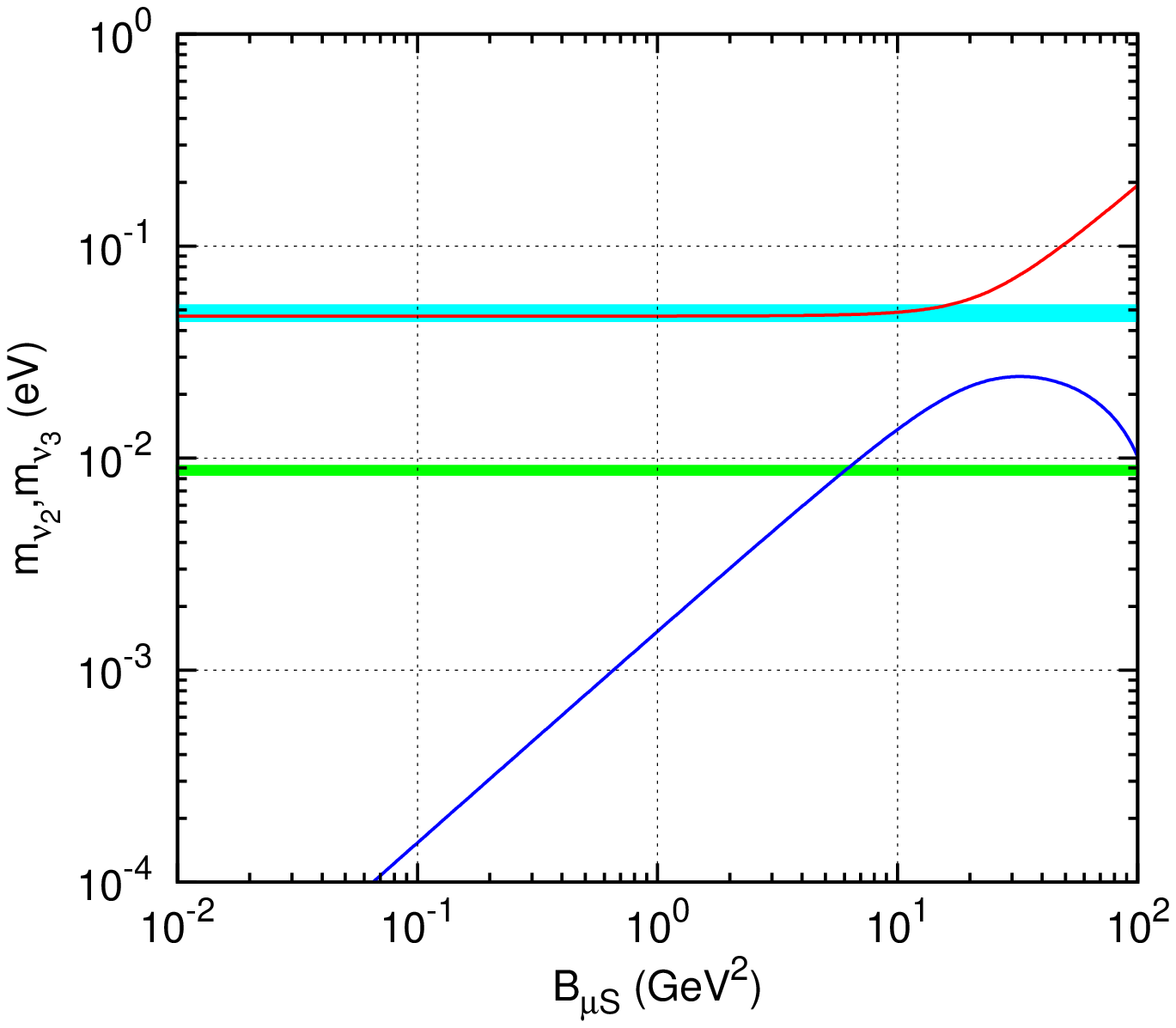}
  \includegraphics[width=0.45\linewidth]{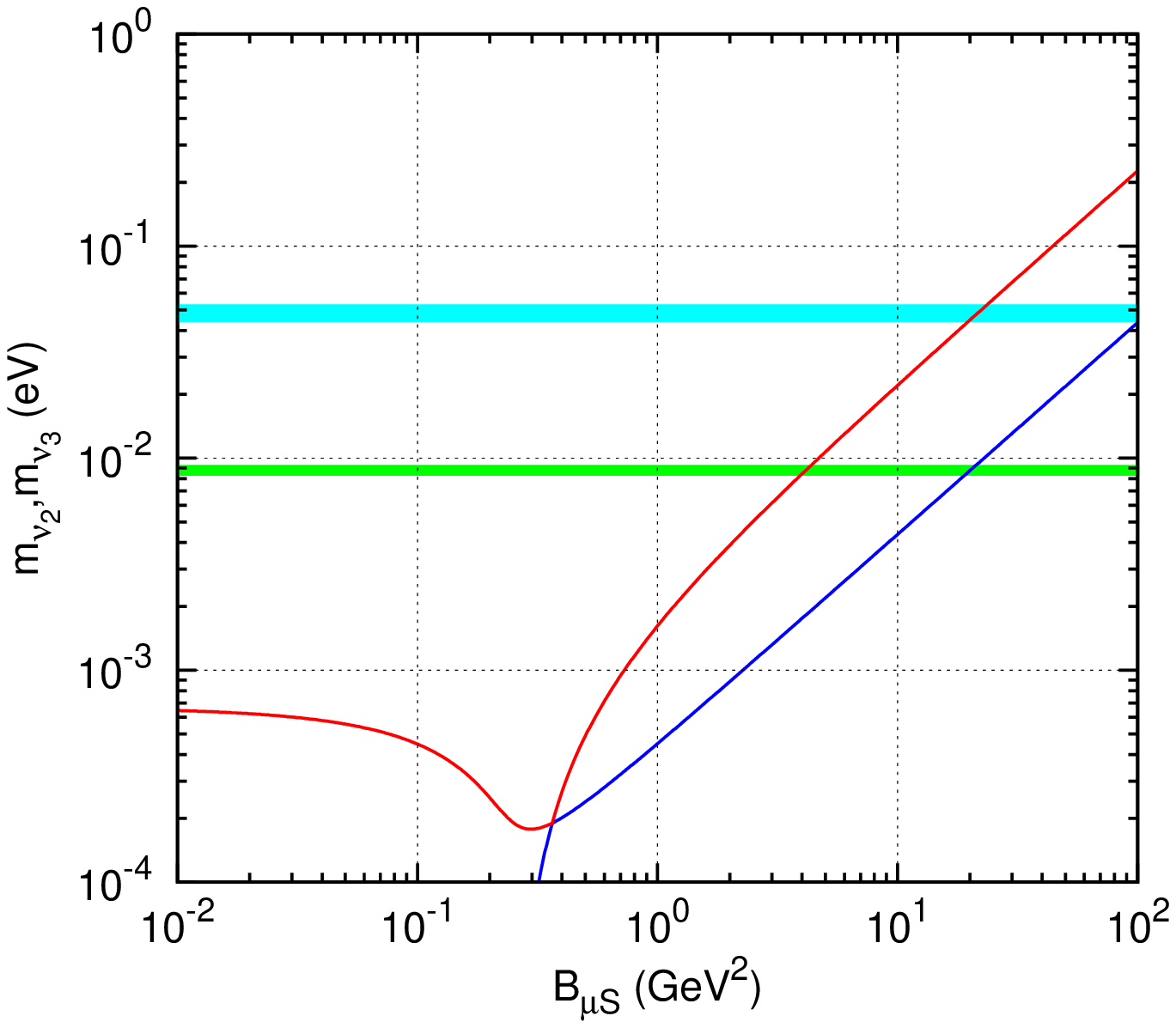}
  \end{tabular}
\caption{Two examples of neutrino mass spectra as a function of 
the parameter $B_{\mu_{S}}$. To the left: 
$\mu_S=7$ eV; to the right $\mu_S=0.1$ eV. 
}
\label{fig:1}
\end{figure}

Neutrino oscillation data require two distinct neutrino mass 
scales, i.e. the atmospheric and the solar scales. Given the 
above discussion, in the MSISM neutrino masses can be fitted 
either with one scale generated by tree-level physics, while 
the other is due to the sneutrino-antisneutrino loop or with  
both scales generated at loop level. An example for each 
case is shown in Fig.~\ref{fig:1}. The left panel shows an 
example for the atmospheric scale being due to tree-level 
physics, with the solar scale generated by loops. The right 
panel shows an example for both masses generated at loop level.  
Which case is realized can not be predicted from the model 
and depends on the relative size of the unknown parameters 
$\mu_S$ and $B_{\mu_S}$. 
Numerical values used in these figures are: $M_2=500$ (GeV),
$\tan\beta=10$, $\mu=150$ (GeV), $M_R=M_S=50$ (GeV),$M_{\nu}=45$
(GeV), $m_{D_1}=0$ (GeV), $-m_{D_2}=m_{D_3}=2.6$ (GeV), $M_{L_1}=700$  (GeV),
$M_{L_2}=750$ (GeV), $M_{L_3}=800$ (GeV), $B_{M_{R}}=50^2 $ (GeV${}^2$), for
$B_{\mu_{S}}=\in[10^{-2},30^2] $ (GeV${}^2$).  Note that on the left
panel $\forall \delta_i =5000$ GeV$^2$, while in the right panel
$\forall \delta_i =1100$ GeV$^2$.
We would like to stress, however, 
that these are just some random examples.

Oscillation data fix two $\Delta m^2$, namely $\Delta m^2_{\rm ATM}$ 
and $\Delta m^2_{\odot}$, but not the absolute scale of neutrino 
masses. Since also the ``sign'' of $\Delta m^2_{\rm ATM}$ is not 
fixed by oscillation data yet, in general three types of spectra 
can fit solar and atmospheric data. These are known in the 
literature as (a) normal hierarchy; (b) inverse hierarchy and 
(c) quasi-degenerate neutrinos. We note that within the MSISM 
it is not possible to get all three light neutrinos degenerate, 
thus we will discuss only (a) and (b).

Any realistic model for neutrino mass must not only explain the 
absolute values for the atmospheric and solar neutrino mass scales, 
but also the corresponding leptonic mixing angles. As first observed 
in \cite{Harrison:2002er}, the so-called tri-bimaximal mixing pattern,
\be{eq:UHPS}
U^{\rm HPS} =
\left(\begin{array}{cccc}
\sqrt{\frac{2}{3}} & \frac{1}{\sqrt{3}} & 0 \cr
- \frac{1}{\sqrt{6}} &  \frac{1}{\sqrt{3}} & - \frac{1}{\sqrt{2}} \cr
- \frac{1}{\sqrt{6}} &  \frac{1}{\sqrt{3}} & \frac{1}{\sqrt{2}}
\end{array}\right) ,
\end{equation}
provides a very good first-order approximation to the measured neutrino 
angles. This pattern can be realized in different ways. However, 
for normal hierarchical neutrinos 
${\cal M}_\nu^{\rm diag}=(0,m_{\odot},M_{\rm ATM})$ 
it leads to the following structure of the neutrino mass matrix in the 
flavour basis:
\be{eq:MHPS}
{\cal M}_{\nu,{\rm NH}}^{\rm HPS} =
\frac{M_{\rm ATM}}{2}\left(\begin{array}{cccc}
0 &  0 &  0 \cr
0 &  1 & -1 \cr
0 & -1 &  1
\end{array}\right) +
\frac{m_{\odot}}{3}\left(\begin{array}{cccc}
1 &  1 &  1 \cr
1 &  1 &  1 \cr
1 &  1 &  1
\end{array}\right).
\end{equation}
Here $M_{\rm ATM}$ ($m_{\odot}$) represents the atmospheric (solar) 
mass scale. 

For the inverse hierarchy,
${\cal M}_\nu^{\rm diag}=(\pm M_{\rm ATM},M_{\rm ATM}+m_S,0)$, 
due to a sign ambiguity in $M_{\rm ATM}$, there are two possible 
textures
\be{eq:MHPS_IH}
{\cal M}_{\nu,{\rm IH1}}^{\rm HPS} =
\frac{M_{\rm ATM}}{2}\left(\begin{array}{cccc}
2 &  0 &  0 \cr
0 &  1 &  1 \cr
0 &  1 &  1
\end{array}\right) +
\frac{m_{\rm S}}{3}\left(\begin{array}{cccc}
1 &  1 &  1 \cr
1 &  1 &  1 \cr
1 &  1 &  1
\end{array}\right),
\end{equation}
\be{eq:MHPS_IH2}
{\cal M}_{\nu,{\rm IH2}}^{\rm HPS} =
\frac{M_{\rm ATM}}{6}\left(\begin{array}{cccc}
-2 &  4 &  4 \cr
4 &  1 &  1 \cr
4 &  1 &  1
\end{array}\right) +
\frac{m_{\rm S}}{3}\left(\begin{array}{cccc}
1 &  1 &  1 \cr
1 &  1 &  1 \cr
1 &  1 &  1
\end{array}\right).
\end{equation}
Here, $m_{\rm S} = \frac{\Delta m^2_{\odot}}{2M_{\rm ATM}}$. Comparing 
eqs.~\rf{eq:MHPS}-\rf{eq:MHPS_IH2} with the index structure of 
eq.~\rf{eq:GenericNeu} 
it is fairly obvious that the MSISM \footnote{And, indeed, any model 
of neutrino mass with this index structure in generation space.} 
can quite easily fit normal hierarchy, whereas the case of inverse 
hierarchy requires a finely tuned cancellation between the different 
contributions (proportional to $\epsilon_i$ and $\delta_i$) to 
eq.~\rf{eq:GenericNeu}. We discuss normal hierarchy first.

Consider the extreme case $b=0$, see eq.~\rf{eq:GenericNeu}. The 
structure required by experimental data could be reproduced with 
$m_{D_1}=0$, $m_{D_2}=-m_{D_3}$ and $\forall \delta_i=\delta$ (and 
vice versa). However, while the relative importance of the terms $a$ 
and $c$ can be independently adjusted by adjusting $B_{M_R}$, $b$ is not 
independent of $a$ and $c$ at the same time. Thus, these equalities 
are not exact. However, one can use this ansatz as a starting point 
and find valid combinations - within the allowed ranges of neutrino 
angles - for $m_{D_i}$ and $\delta_i$ by a simple iterative procedure. 
In the numerical scans shown in the next section we have always allowed 
that the range of $m_{D_i}/m_{D_j}$ and $\delta_i/\delta_j$ vary 
randomly within some moderate factor such that all of the allowed 
range in the neutrino angles are covered. With such a random 
selection of parameters we can fit all angles easily, however, 
there is no prediction and no ``typical'' size of any neutrino 
angle.

The case of inverse hierarchy, however, requires that the tree-level 
and 1-loop contribution to the neutrino mass matrix are finely tuned 
against each other. If, for example, we compare the first texture 
for inverse hierarchy eq.~\rf{eq:MHPS_IH} with eq.~\rf{eq:GenericNeu} 
one finds
\baq{eq:finetune}
a \epsilon_1^2 + 2 b \epsilon_1 \delta_1 + c \delta_1^2 & =&  M_{\rm ATM} 
\\
a \epsilon_1\epsilon_2 + b (\epsilon_1 \delta_2 + \epsilon_1 \delta_1)  
+ c \delta_1\delta_2 & =&   \frac{m_{\rm S}}{3}
\\
a \epsilon_2^2 + 2 b \epsilon_2 \delta_2 + c \delta_2^2 & =&  
\frac{M_{\rm ATM}}{2}
\eaq
i.e. tree-level and 1-loop contributions have to be tuned to cancel 
each other up to $\frac{m_{\rm S}}{M_{\rm ATM}}$ in order to reproduce 
the desired texture. Similar relations hold for the other texture, 
eq.~\rf{eq:MHPS_IH2}. We did not attempt to find such fine-tuned solution 
in the numerical scans discussed in the next section.

%-----------------------------------------------------------------------------
\section{Lepton flavour violation and collider signals}
%-----------------------------------------------------------------------------
\label{sec:pheno}

In this section we discuss phenomenological aspects of the MSISM. 
We will concentrate on LFV charged lepton 
decays and the decays of charginos to charged leptons and singlet 
sneutrinos. In general, the new singlets of the MSISM could appear in 
decay chains at the LHC if either (or both) $m_{D}$ or $\delta$ are 
large, as expected in the MSISM, thus potentially altering the 
phenomenology with respect to MSSM expectations. However, the probably 
most interesting part of the parameter space is that where one of 
the scalar singlets is the LSP, thus being potentially a DM candidate. 
In this case scalar singlets are guaranteed to show up at the end of 
the supersymmetric decay chains. We will exclusively concentrate on 
this case in our discussion of chargino decays below. Note, however, 
that LFV $\ell_j\to \ell_i \gamma$ decays are independent of this assumption.

We consider the decays
\be{eq:decay}
\widetilde\chi^\pm_1\to\widetilde N_a +\ell^\pm_i~,\qquad
a=1,\dots,4~,\quad\ell_i=e,\mu,\tau~,
\ee
with
$\widetilde N_1$ ($\widetilde N_3$) being the CP conjugated state to 
$\widetilde N_2$ ($\widetilde N_4$). 
The relevant piece of the Lagrangian for the calculation of 
the decay widths of \rf{eq:decay} is
\be{eq:LangCS}
{\mathcal L}_{\ell\chi^-\tilde\nu}=\bar{\widetilde\chi}^-_j(C^L_{ija} 
P_L+C^R_{ija} P_R)
\ell_i \widetilde N_a+{\rm h.c.}~,
\ee
with
\baq{eq:CoupCS}
C^R_{ija}&=&\frac{1}{\sqrt{2}} Y_{\ell_i}U_{j2}(G_{ai}-iG_{a(i+5)})~,
\nonumber\\[2mm]
C^L_{ija}&=&-\frac{1}{\sqrt{2}}\left[
g V_{j1}^* (G_{ai}-iG_{a(i+5)})-h_\nu^i V_{j2}^* (G_{a4}-iG_{a9})
\right]~,
\eaq
where the charged lepton Yukawa couplings are 
$Y_{\ell_i}=\frac{g}{\sqrt{2}} \frac{m_{\ell_i}}{m_W \cos\beta}$, 
$\ell_i=e,\mu,\tau$, and
$U$ and $V$ are the unitary 2$\times2$ chargino mixing matrices, 
which diagonalize the chargino mass matrix by
$U^* M_{\chi^\pm} V^{-1}={\rm diag}(m_{\chi^\pm_1},m_{\chi^\pm_2})$, 
with $m_{\chi^\pm_k}>0$. 
The decay widths of the decays \rf{eq:decay} are finally given as
\be{eq:WidthC}
\Gamma(\widetilde\chi^\pm_1\to\widetilde N_a +\ell^\pm_i)=
\frac{(m_{\chi^\pm_1}^2-m_{\tilde N_a}^2)^2}{32\pi~ m_{\chi^\pm_1}^3}
\left(|C^L_{i1a}|^2+|C^R_{i1a}|^2\right)~.
\ee
As the members of each CP conjugated pair are always nearly degenerate, $m_{\tilde N_1}\approx m_{\tilde N_2}$ and $m_{\tilde N_3}\approx m_{\tilde N_4}$, they (most likely) cannot be distinguished experimentally. For this reason, we sum over the CP-even and associated CP-odd sneutrino states of each CP conjugated pair
\baq{eq:DefSumWidthC}
\Gamma(\widetilde\chi^\pm_1\to\widetilde N_{1+2} +\ell^\pm_i) &\equiv&
\Gamma(\widetilde\chi^\pm_1\to\widetilde N_1 +\ell^\pm_i) + \Gamma(\widetilde\chi^\pm_1\to\widetilde N_2 +\ell^\pm_i)~,\nonumber\\
\Gamma(\widetilde\chi^\pm_1\to\widetilde N_{3+4} +\ell^\pm_i) &\equiv&
\Gamma(\widetilde\chi^\pm_1\to\widetilde N_3 +\ell^\pm_i) + \Gamma(\widetilde\chi^\pm_1\to\widetilde N_4 +\ell^\pm_i)~.
\eaq
To understand the dependence of the decay widths in eq.~\rf{eq:WidthC}
on the model parameters, one can use an approximate diagonalization 
of the sneutrino sector as discussed above. If $M_{\tilde L}, M_R
\gg\varepsilon_i,\delta_i$, the leading contribution to the decay width 
to the lightest CP conjugated pair, 
$\Gamma(\widetilde\chi^\pm_1\to\widetilde N_{1+2} +\ell^\pm_i)$, 
according to eq.~\rf{eq:WidthC} is given by 
\baq{eq:approx}
\sum_{a=1}^2|C^L_{i1a}|^2 \approx 
\varepsilon_i^2 ~\frac{|V_{12}|^2}{2 M_R^2 v_u^2} 
(\cos^2\theta_++\cos^2\theta_-)~, & {\rm if} & 
 \varepsilon_i \gg \delta_i
\nonumber\\ 
\sum_{a=1}^2|C^L_{i1a}|^2\approx
\delta_i^2~\frac{g^2|V_{11}|^2}{2\widehat M_L^4}
(\cos^2\theta_++\cos^2\theta_-)~, & {\rm if} &
 \varepsilon_i \ll \delta_i
\eaq
The results for the decays into the heavier second pair of singlet 
sneutrino states, 
$\Gamma(\widetilde\chi^\pm_1\to\widetilde N_{3+4} +\ell^\pm_i)$, 
are obtained by replacing
$(\cos^2\theta_++\cos^2\theta_-)\to (\sin^2\theta_++\sin^2\theta_-)$
in \rf{eq:approx}. 

In our numerical calculations, we have fixed the parameters as follows:
$M_2=700$ GeV, $\tan\beta=5$, $\mu=400$ GeV, $M_{L_i}=700$ GeV,
$M_R=M_S=M_{\nu}=200$ GeV, $B_{M_R}=(200)^2$ GeV$^2$. This choice is 
motivated by eq.~\rf{eq:approx} which shows that the higgsino 
component of the chargino couples proportional to $\epsilon_i^2$ to 
charged leptons. Other parameters have been randomly generated: 
$(\sum_i m_{D_i}^2)^{1/2}\in 10^{[-4,\,2.6]}$, 
$(\sum_i \delta_i^2)^{1/4}\in 10^{[-4,\,3]}$. Neutrino data on mixing 
angles (and mass scales) constrains the other parameters. In the numerical 
examples we adjust the parameters $\mu_S$ and $B_{\mu_S}$ in such 
a way that the atmospheric neutrino mass scale is determined by the tree-level 
neutrino mass matrix contribution, eq.~\rf{eq:SeeSaw}, 
while the solar neutrino mass scale is 
obtained by the 1-loop correction. The component $m_{D_1}$ then has to 
be considerably smaller than the components $m_{D_2} \sim m_{D_3}$, 
so that the reactor neutrino angle is small and the atmospheric
neutrino mixing angle is maximal; the components $\delta_i$ are all of
the same order so that the solar mixing angle is large. Note that we 
have imposed neutrino data to be in agreement with the experimental 
$3\sigma$ allowed range. Also note that in all the plots 
%(except when explicitly pointed out) 
we have imposed the experimental upper bounds 
on the low energy LFV radiative decays BR($\ell_j\to\ell_i + \gamma$).

In order to quantify whether the main contribution to the chargino 
decays is due to the parameters $m_{D_i}$ or the parameters $\delta_i$, 
in our numerical analysis we define the ratio 
\be{eq:defr}
r\equiv\frac{(\sum_i m_{D_i}^2)^{1/2}}{(\sum_i \delta_i^2)^{1/4}}.
\ee
We will concentrate on the case where $m_{D_i}$ gives the dominant 
contribution to the chargino decay ($r>1$).  Some comments on the 
other extreme are given near the end of this section. 

%-----------------------------------------------------------------------------
%\subsection{$r>1$}
%-----------------------------------------------------------------------------

\begin{figure}[htb]
  \centering
  \begin{tabular}{ccc}
    \includegraphics[width=0.48\textwidth]{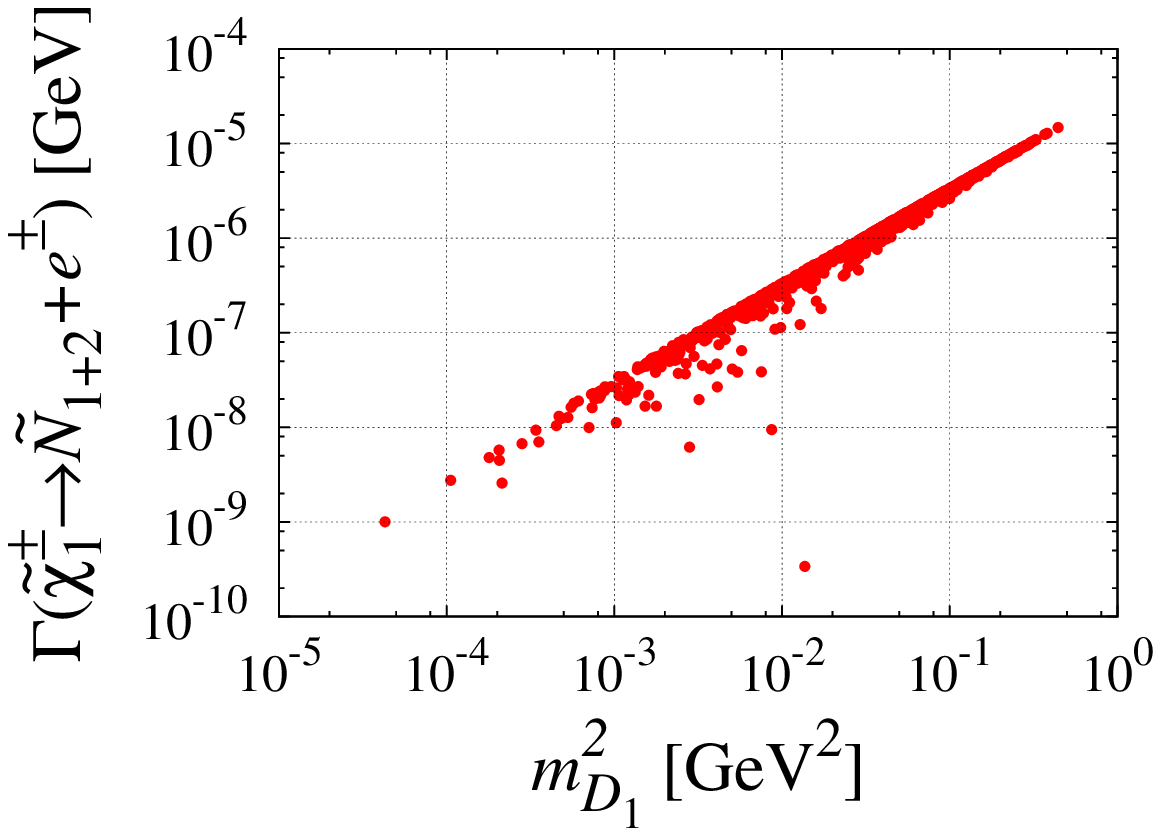}&
    \includegraphics[width=0.48\textwidth]{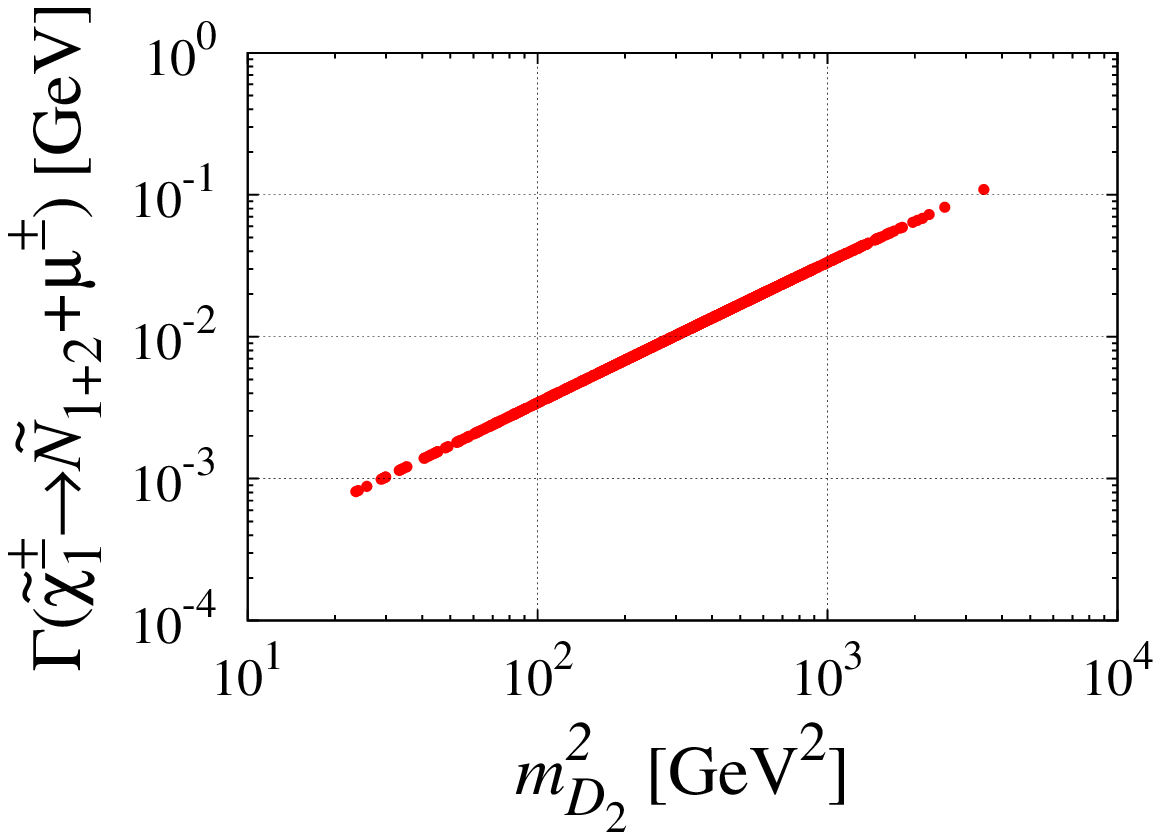}
  \end{tabular}
  \caption{Decay width $\Gamma(\widetilde\chi^\pm_1\to \widetilde N_{1+2} + \ell^\pm_i)$ 
of the lightest chargino to the lightest pair of quasi-degenerate CP conjugated sneutrinos 
and a charged lepton ($\ell_i=e$ in the left panel and $\ell_i=\mu$ in the right panel) 
as a function of the parameter $m^2_{D_1}$ (left panel) and $m^2_{D_2}$ (right panel). 
The plot for $m^2_{D_3}$ is very similar to the one for $m^2_{D_2}$ and thus not 
shown. 
All plots correspond to the case $r>1$, see eq.~\rf{eq:defr}.
}
  \label{fig:Gamma-mD-rLarge}
\end{figure}

For the case $r>1$, Fig.~\ref{fig:Gamma-mD-rLarge}  shows the correlation of 
the decay width of the lightest chargino to the lightest pair of quasi-degenerate 
CP conjugated sneutrinos ($\widetilde N_1$ and $\widetilde N_2$) and a 
charged lepton $\ell_i$ with respect to the corresponding parameter $m^2_{D_i}$. 
We have checked that this correlation also holds for the chargino decay width 
$\Gamma(\widetilde\chi^\pm_1\to \widetilde N_{3+4} + \ell^\pm_i)$, which involves
the second lightest pair of quasi-degenerate CP conjugated sneutrinos 
($\widetilde N_3$ and $\widetilde N_4$). 
This behaviour is as expected  from the analytical approximation in eq.~\rf{eq:approx}.
Note, however, that the correlation of the widths involving the electron 
with respect $m^2_{D_1}$ are not as clean than the others. This is due to the 
constraint on the neutrino reactor angle imposed by neutrino data, which 
requires $m_{D_1}$ to be much smaller than $m_{D_2}$ and $m_{D_3}$. 
Comparing the size of these calculated widths to typical widths for final 
states $\chi^\pm_1\to \chi^0 + W^\pm$ and $\chi^\pm_1\to \chi^0 + \ell^\pm\nu$
one finds that branching ratios into muon and tau final states can be sizeable, 
whereas the width to final state $\widetilde N_{1+2} + e^\pm$ is expected to be 
too small to be measurable.

%%%%%%%%%%%%%%%%%%%%%%%%%%%%%%%%%

We note in passing that the product of the decay widths of the
lightest chargino to one the two lightest pairs 
of quasi-degenerate CP conjugated sneutrinos and a charged
lepton $\ell_i$ times the same width but to the charged lepton
$\ell_j$ are correlated with the low energy LFV process $\BR(\ell_j\to
\ell_i\gamma)$. Again, the correlation involving the electron in the
final state is less strong than the ones involving only $\mu$ and
$\tau$ because of the relative smallness of the parameter $m_{D_1}$
imposed by the experimental upper bound on the neutrino reactor
angle. Since the absolute widths, however, will not be measurable 
at the LHC, more interesting phenomenologically are ratios of partial 
widths, i.e. ratios of branching ratios.

\begin{figure}[htb]
  \centering
  \begin{tabular}{cc}
    \includegraphics[width=0.46\textwidth]{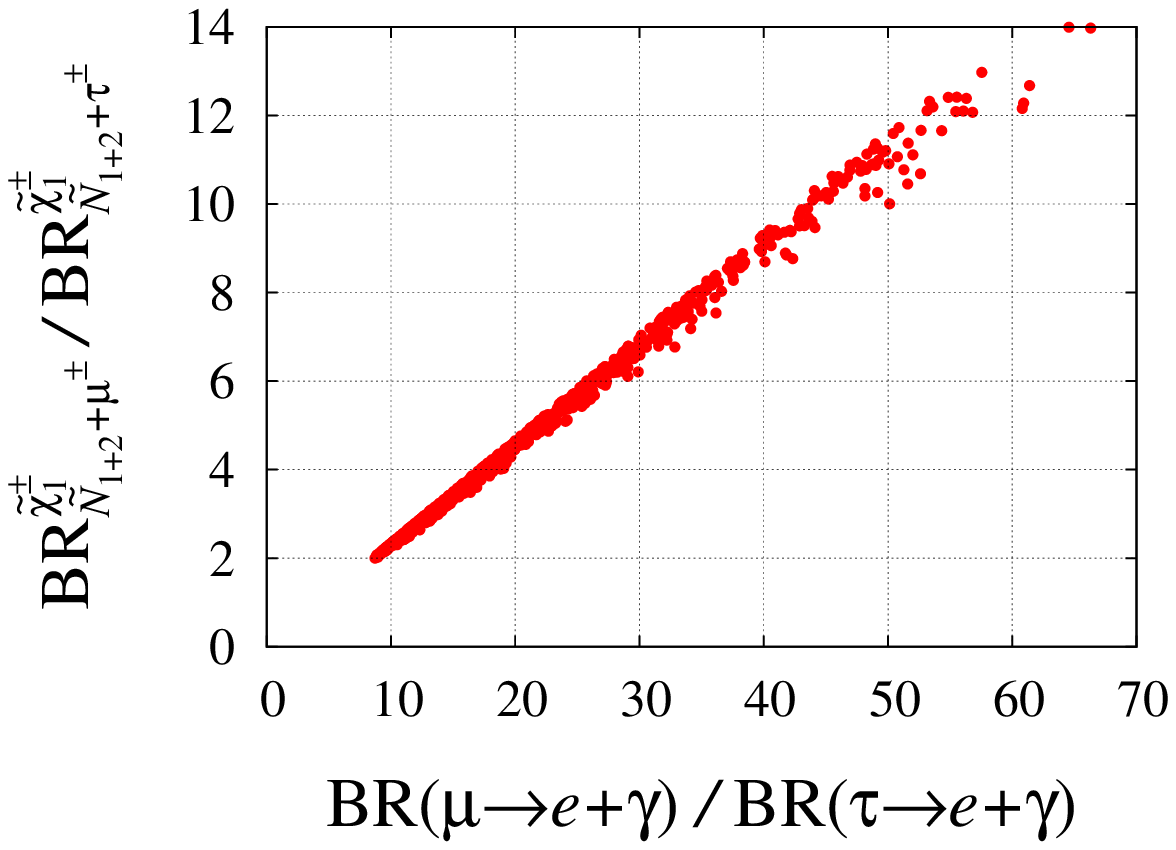}&
    \includegraphics[width=0.46\textwidth]{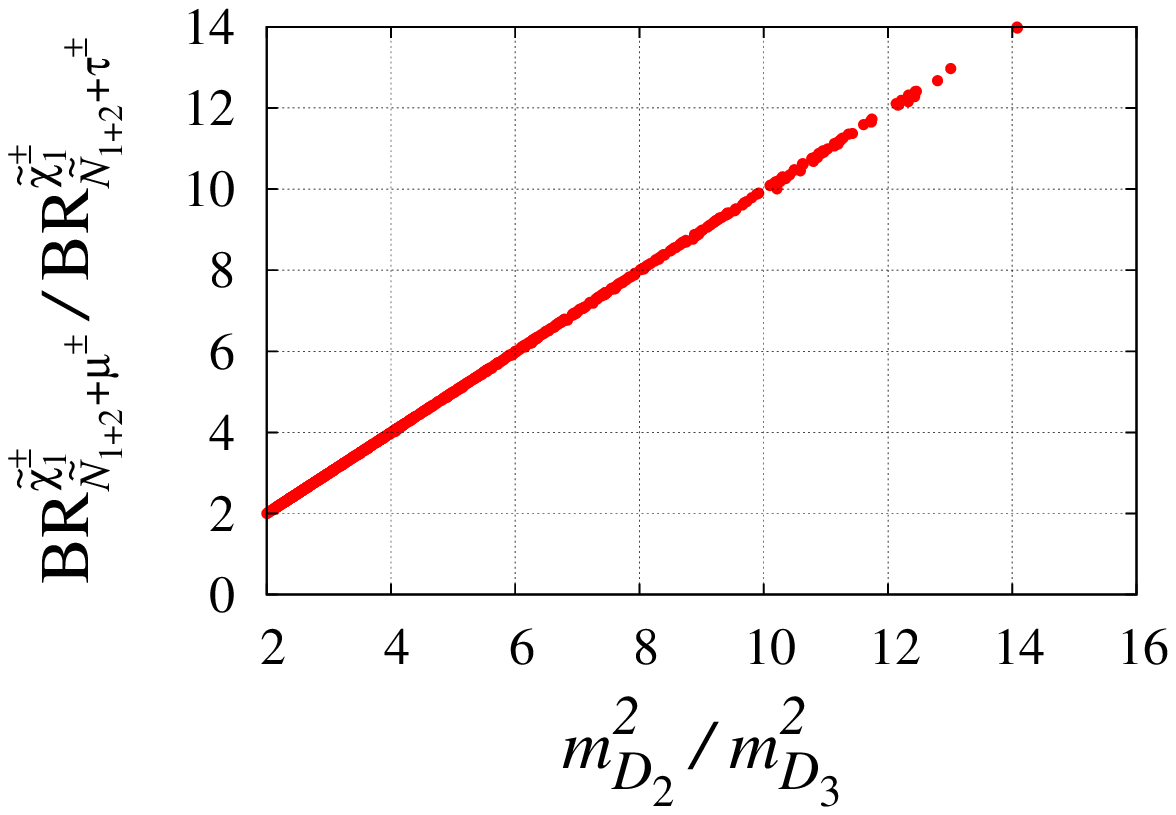}
  \end{tabular}
  \caption{Ratio of branching ratios $\BR(\widetilde\chi^\pm_1\to \widetilde
N_{1+2} + \mu^\pm)/\BR(\widetilde\chi^\pm_1\to \widetilde N_{1+2} +\tau^\pm)$ 
as a function of $\BR(\mu \to e + \gamma)/\BR(\tau \to e \gamma)$ (left panel) 
and $m^2_{D_2}/m^2_{D_3}$ (right panel).}
  \label{fig:ratofrat}
\end{figure}

Fig.~\ref{fig:ratofrat} shows ratios of branching ratios 
$\BR(\widetilde\chi^\pm_1\to \widetilde N_{1+2} + \mu^\pm)/
\BR(\widetilde\chi^\pm_1\to \widetilde N_{1+2} +\tau^\pm)$ 
as a function of $\BR(\mu \to e + \gamma)/\BR(\tau \to e \gamma)$ (left panel) 
and $m^2_{D_2}/m^2_{D_3}$ (right panel). Again, the same correlations can be found for 
$\BR(\widetilde\chi^\pm_1\to \widetilde N_{3+4} + \mu^\pm)/
\BR(\widetilde\chi^\pm_1\to \widetilde N_{3+4} +\tau^\pm)$. 
A measurement of both, chargino decays 
and LFV lepton decays, would therefore constitute a consistency check of the 
scenario we discuss. Note, however, that the expected branching ratio for 
$\BR(\tau \to e \gamma)$ is quite small (at most $10^{-12}$) 
compared to current experimental sensitivities. 

%%%%%%%%%%%%%%%%%%%%%%%%%%%%%%%%%

\begin{figure}[htb]
  \centering
  \begin{tabular}{cc}
    \includegraphics[width=0.46\textwidth]{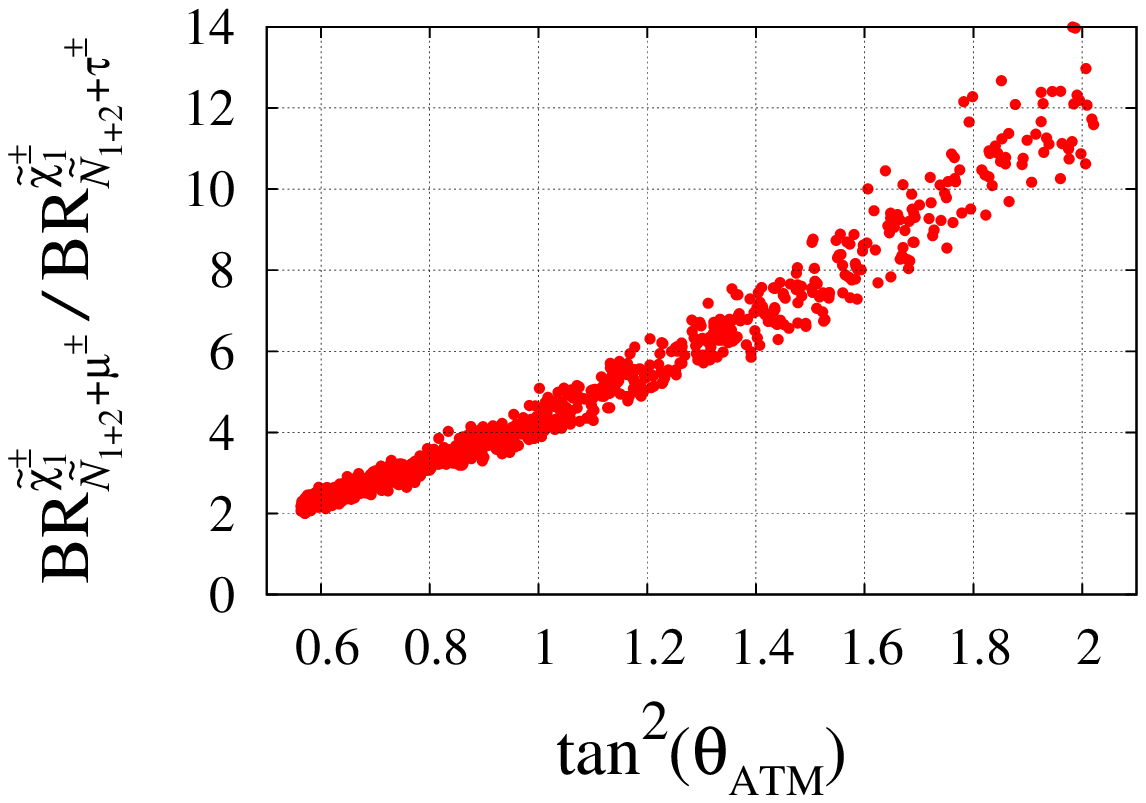}&
    \includegraphics[width=0.46\textwidth]{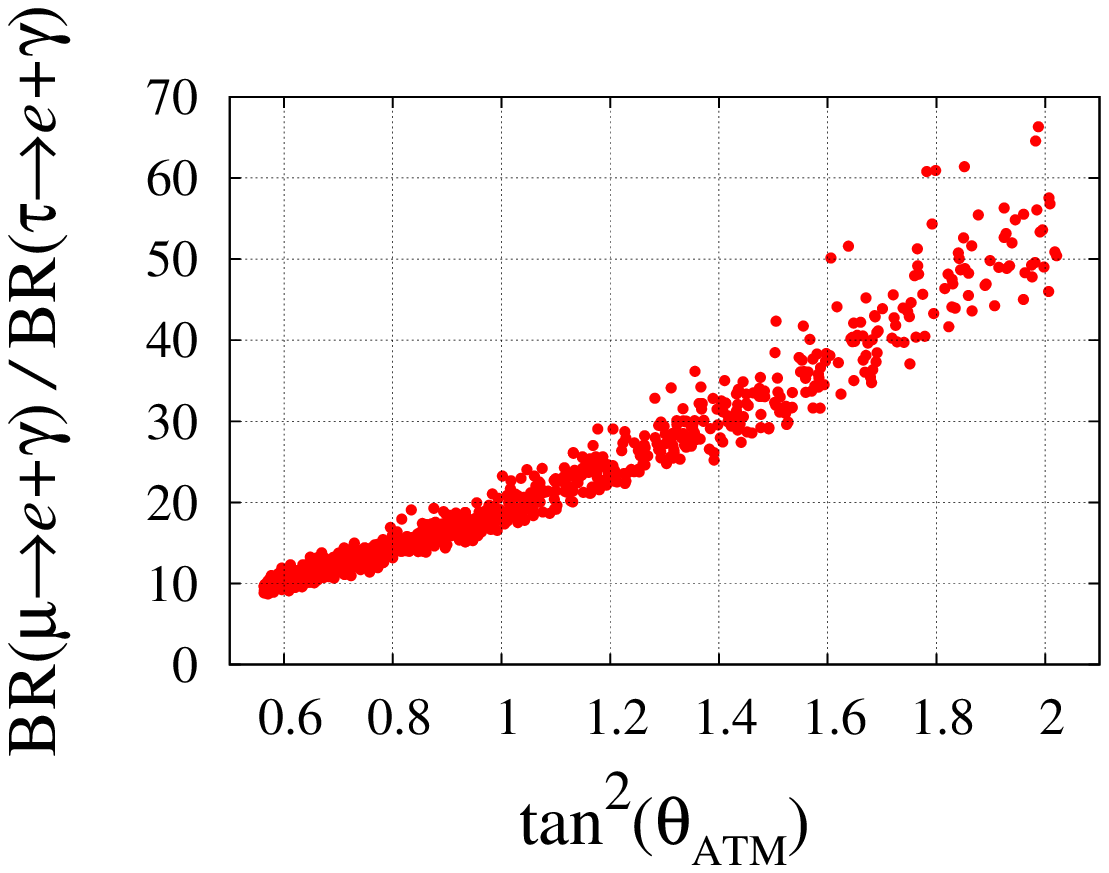}
  \end{tabular}
  \caption{In the left panel, ratio of branching ratios 
$\BR(\widetilde\chi^\pm_1\to \widetilde
N_{1+2} + \mu^\pm)/\BR(\widetilde\chi^\pm_1\to \widetilde N_{1+2} +
\tau^\pm)$ as a function of the atmospheric neutrino mixing angle,
$\tan^2(\theta_{\rm ATM})$.  In the right panel, 
$\BR(\mu \to e + \gamma)/\BR(\tau \to e \gamma)$ as a function of 
$\tan^2(\theta_{\rm ATM})$. 
Both plots correspond to the case of $r>1$.}
  \label{fig:Gamma-ATMangle-rLarge}
\end{figure}

Left panel in Fig.~\ref{fig:Gamma-ATMangle-rLarge} shows the correlation of the ratio 
of branching ratios of the lightest chargino decaying to the lightest sneutrino pair 
and $\mu$ divided by its decay to the lightest sneutrino pair and $\tau$ as a function 
of the atmospheric neutrino mixing angle. Recall that these data points have parameters 
chosen such that the atmospheric scale is generated by tree-level physics. The correlation 
exists for the lightest and for the next-to-lightest pair of singlet sneutrinos, 
if kinematically accessible.
Right panel in Fig.~\ref{fig:Gamma-ATMangle-rLarge} shows the correlation of the ratio 
of $\BR(\mu \to e + \gamma)$ divided by  $\BR(\tau \to e + \gamma)$ as a function of the 
atmospheric neutrino mixing angle. 
As can be seen from both panels in Fig.~\ref{fig:Gamma-ATMangle-rLarge}, 
the neutrino sector (the atmospheric mixing angle) is related to 
collider observables (the LFV decays of the lightest chargino to a singlet sneutrino and a lepton) 
as well as low energy LFV observables (the radiative decays of the charged leptons). 

%------------------------------------------------------------
% g-2
%------------------------------------------------------------

As in our model we have relatively light right handed neutrinos and
sneutrinos with large Yukawa coupling we should check the
contributions from these new particles to the muon $g-2$. We have
calculated the new contributions to $a_{\mu}$ and verified that 
 our numerical points - once they pass the cuts from $l_i \to l_j +
 \gamma$ - also pass the experimental constraint from
 $a_{\mu}$~\cite{Amsler:2008zzb}.

%-----------------------------------------------------------------------------
%\subsection{$r<1$}
%-----------------------------------------------------------------------------
Finally we would like to comment on the case $r \ll 1$, i.e. the parameters 
$\delta_i$ giving the dominant contribution to the neutrino mass matrix 
and thus to the lightest chargino LFV decays. 
We have scanned the parameter space of the model for such solutions and, 
as expected the decays of the charginos to singlet sneutrinos plus a lepton correlate 
with $\delta_i$ instead of $m_{D_i}$ in this extreme. However, in all 
points we have found the absolute widths for the final states $\widetilde N_a$ % + \ell_i^{\pm}$ 
plus a charged lepton are much smaller than in the case $r> 1$ discussed above  
(at most of the order of $10^{-5}$ GeV). 
One expects therefore that the corresponding branching ratios are too small 
to be measured at LHC.

%----------------------------------------------------------------------------
\section{Summary \label{conclusion}}
%----------------------------------------------------------------------------

The minimal supersymmetric inverse seesaw model (MSISM) with only one 
pair of singlet superfields can explain all existing neutrino 
oscillation data. We have calculated the neutrino mass matrix 
at 1-loop order and discussed the constraints on model parameters 
due to the experimentally measured leptonic mixing angles and 
neutrino masses.

Since in the MSISM one expects the new singlet fields to exist at 
a mass scale below (approximately) TeV, additional phenomenology 
is expected to show up in experiments searching for lepton flavour 
violation, such as $\mu\to e \gamma$, and possibly at the LHC. 
Absolute values of branching ratios can not be predicted, but the 
minimal model relates the observed atmospheric angle to some 
specific ratios of branching ratios.  A measurement of these ratios 
can therefore potentially serve as a test of our minimal model.

For the LHC we have concentrated in our discussion on the case that one 
of the singlet scalar fields, a mixture of the scalar neutrino and the 
scalar singlet $\widetilde S$, is the lightest supersymmetric particle. This 
assumption is motivated by the observation that this singlet could be 
the CDM. Charginos can then decay to charged leptons plus singlet 
sneutrinos. A measurement of these decays and low energy lepton flavour 
violating lepton decays, such as $\mu\to e +\gamma$ and $\tau \to e +\gamma$ 
would provide an interesting test of the minimal supersymmetric inverse 
seesaw model.

%----------------------------------------------------------------------------
\section*{Acknowledgments}
%----------------------------------------------------------------------------

This work was partially supported by FCT through the projects
CFTP-FCT UNIT 777 and CERN/FP/83503/2008, which are partially
funded through POCTI (FEDER), and by the Marie Curie RTN MRT-CT-2006-035505. 
M.H. is supported by the Spanish grant FPA2008-00319/FPA.
A.~V.~M. is supported by {\it Funda\c c\~ao para a Ci\^encia 
e a Tecnologia} under the grant SFRH/BPD/30450/2006.

%----------------------------------------------------------------------------
\appendix
\appendixpage
\renewcommand{\theequation}{\Alph{section}.\arabic{equation}}
%----------------------------------------------------------------------------

%-----------------------------------------------------------------------------
\section{Higgs-heavy neutrino loop}
%-----------------------------------------------------------------------------
\setcounter{equation}{0}

Here we consider the 1-loop contributions to the neutrino mass
matrix which are mediated through the Higgs-heavy neutrino loops.
For their calculation, the required Lagrangian is given by
\be{eq:Lagrange2}
{\mathcal L}_{H^0\nu\nu^L}=-h_\nu^i E_{(k+3)4}^* 
U^{\rm tr}_{im}\bar{\nu}^L_k P_L \nu_m H_u^0+{\rm h.c.}~,
\ee
where $E$ denotes the unitary mixing matrix, which diagonalizes 
the mass matrix of the neutral fermion fields of eq.~\rf{eq:Neutrino},
$E^*M^\nu E^{-1}={\rm diag}(m_{N_i})
={\rm diag}(0,0,m_{\nu_3},m_{L_1},m_{L_2})$.
We note that only the contribution of the two heavy neutrinos is 
significant, and thus in \rf{eq:Lagrange2} $\bar{\nu}^L_k$, $k=1,2$, 
denotes the two heavy neutrinos which are a mixture of the fermionic 
states $\nu^c$ and $S$. Contributions of light neutrinos in the loop 
are negligible, since the corresponding mixing elements $E_{i4}$, 
$i=1,2,3$, are tiny. The neutral component of the up-type Higgs doublet 
can be expressed in terms of mass eigenstates as follows \cite{Gunion:1984yn}
\be{eq:Higgsfield}
H^0_u=v_u+\frac{1}{\sqrt{2}}
\left[
\cos\alpha h^0+\sin\alpha H^0+i(\cos\beta A^0+\sin\beta G^0)
\right]~,
\ee
where $h^0,H^0$ are the two CP-even scalar fields, with $m_{h^0}<m_{H^0}$ 
and the corresponding mixing angle $\alpha$, $A^0$ is the CP-odd scalar 
field, while $G^0$ is the Goldstone field, with $\tan\beta\equiv v_u/v_d$.
From the calculation of the self-energy functions of these contributions 
we obtain
\begin{eqnarray}
\!\!\!\!\!\!\!\!\!\Sigma^{mn}_{S1} &=& \frac{-m_{L_k}}{2(4\pi)^2}
{\mathcal H}_r^2 h_\nu^i h_\nu^j U^{\rm tr}_{im} U^{\rm tr}_{jn}E_{(k+3)4}^{*2}
 B_0(m^2_{L_k},m^2_{{\mathcal H}_r})~,\label{eq:AmplS1}\\[2mm]
\!\!\!\!\!\!\!\!\!\Sigma^{mn}_{V1} &=& \frac{-1}{2(4\pi)^2}
{\mathcal H}_r^2 h_\nu^{i*} h_\nu^j U^{\rm tr*}_{im} U^{\rm tr}_{jn} 
|E_{(k+3)4}|^2 
 B_1(m^2_{L_k},m^2_{{\mathcal H}_r})~,
\label{eq:AmplV1}
\end{eqnarray}
where we have introduced the shorthand notations
${\mathcal H}_r=(\cos\alpha,\sin\alpha,i\cos\beta,i\sin\beta)$ and
$m_{{\mathcal H}_r}=(m_{h^0},m_{H^0},m_{A^0},m_Z)$, to sum up 
the various Higgs boson contributions. 
In eqs.~(\ref{eq:AmplS1})-(\ref{eq:AmplV1}), 
the standard 2-point loop integrals $B_0(x,y)$ and $B_1(x,y)$, 
when evaluated at zero momentum ($p^2=0$), can be written as
\baq{eq:Loopfuct}
B_0(x,y) &=&\Delta+1+\log\frac{Q^2}{y}-\frac{x}{x-y}\log\frac{x}{y}~,\\[2mm]
B_1(x,y) &=&-\frac{1}{2}\left[
\Delta+1+\log\frac{Q^2}{y}-\frac{x}{x-y}\log\frac{x}{y}
\right].
\eaq
We see from eqs.~(\ref{eq:AmplS1})-(\ref{eq:AmplV1}) that the flavor structure
is determined by the product of the neutrino Yukawa couplings 
$h_\nu^m h_\nu^n$ (in the flavor basis), and has therefore the same structure 
as the tree-level contribution, see eq.~\rf{eq:SeeSaw}.
As a result, only the 33-element in $\Sigma^{mn}_{S1}$ and 
$\Sigma^{mn}_{V1}$ receive non-vanishing contributions.
Thus, by including only the Higgs-heavy neutrino loop contributions
still two of the light neutrinos remain massless.

%-----------------------------------------------------------------------------
\section{LFV lepton decays}
%-----------------------------------------------------------------------------
\setcounter{equation}{0}

Here we summarize the formulas for the calculation of the two-body LFV 
lepton decay rates in the MSISM with only one generation of singlet superfields.
The formulas are derived from the superpotential in eq.~\rf{eq:SuperPot}
and the soft SUSY breaking Lagrangian in eq.~\rf{eq:softSUSY}.  
In the context of the SUSY inverse seesaw mechanism with three generation of 
singlet superfields and mSugra boundary conditions see Ref.~\cite{Deppisch:2004fa}.  

The gauge invariant amplitudes of the decays 
$\ell^-_j(p)\to \ell^-_i(p-q)+\gamma(q)$,
$\ell_j=\mu,\tau; \ell_i=e,\mu$, can be defined as \cite{Lavoura:2003xp}
\be{eq:Amplrare}
T=i e \epsilon_{\mu}^*(q)\bar u_{\ell_i}(p-q)[\sigma^{\mu\nu}q_\nu
(\sigma_{L,ij} P_L+\sigma_{R,ij} P_R)]
u_{\ell_j}(p)~.
\ee
In the calculation of the left and right amplitudes, $\sigma_{L,R}$,
we neglected terms proportional to the small lepton mass $m_{\ell_i}$.
The heavy lepton contributions 
give rise to the right amplitude as \cite{Lavoura:2003xp,Ilakovac:1994kj}
\be{eq:RHL}
\sigma^{\rm HL}_{R,ij}=\frac{ig^2}{32\pi^2m_W^2}m_{\ell_j}
\sum_{k=1}^5~E^*_{kj}E_{ki}~\left(
\frac{-4s^3+45s^2-33s+10}{4(s-1)^3}-\frac{3s^3}{2(s-1)^4}\ln s\right)
\ee
with $s=m_{N_k}^2/m_W^2$. 
Note that the loop function in
eq.~\rf{eq:RHL} coincides with eq.~(68) of \cite{Lavoura:2003xp}, and
differs by a constant, $-5/6$, from eq.~(B.2) of
Ref.~\cite{Ilakovac:1994kj}. For $\ell^-_j(p)\to
\ell^-_i(p-q)+\gamma(q)$ decays this constant does not contibute, due
to the unitarity of the coupling matrices, but for $g-2$ the main
contribution for light neutrinos comes precisely from this constant
and therefore the correct loop function for both cases is
eq.~\rf{eq:RHL}. 
The sneutrino-chargino loop contributions to the right amplitude
read
\baq{eq:RS}
\sigma^{\rm SC}_{R,ij}=-\frac{i}{16\pi^2}\sum_{k=1}^2\sum_{a=1}^{10}&&
\left[C_{ika}^{L*}C_{jka}^L \frac{m_{\ell_j}}{m^2_{\tilde N_a}}
\left(\frac{t^2-5t-2}{12(t-1)^3}+\frac{t\ln t}{2(t-1)^4}\right)\right.
\nonumber\\[2mm]
{}&&\left.+ C_{ika}^{L*}C_{jka}^R\frac{m_{\chi^-_k}}{m^2_{\tilde N_a}}
\left(\frac{t-3}{2(t-1)^2}+\frac{\ln t}{(t-1)^3}\right)
\right]~,
\eaq
with $t=m^2_{\chi^\pm_k}/m^2_{\tilde N_a}$ and the couplings 
defined in eq.~\rf{eq:CoupCS}. The sneutrino-chargino loop contributions 
to the left amplitude are obtained from the right ones by interchanging the 
left and right chiral couplings, i.e. 
$\sigma^{\rm SC}_{L,ij}=\sigma^{\rm SC}_{R,ij}(L\leftrightarrow R)$.
With the definition of the amplitude in eq.~\rf{eq:Amplrare},
the corresponding decay widths are given by
\be{eq:DecayW}
\Gamma(\ell_j\to\ell_i \gamma)
=\frac{\alpha}{4} m_{\ell_j}^3(|\sigma^{\rm SC}_{L,ij}|^2
+|\sigma^{\rm SC}_{R,ij}+\sigma^{\rm HL}_{R,ij}|^2)~,
\ee
with $\alpha=e^2/4\pi$ and where again in the kinematics we have neglected 
the terms proportional to the small lepton mass $m_{\ell_i}$.

%-----------------------------------------------------------------------------
\section{Muon anomalous magnetic moment}
%-----------------------------------------------------------------------------
\setcounter{equation}{0}

The formulas for the muon anomalous magnetic moment can be derived
from those in the previous section. Defining, as usual,
\begin{equation}
  \label{eq:C1}
  a_{\mu} = \frac{g-2}{2}
\end{equation}
we have from eq.~\rf{eq:Amplrare}
\begin{equation}
  \label{eq:C2}
  a_{\mu} = -2\, m_{\mu}\ \frac{1}{2} \left(\sigma_{L,22}+\sigma_{R,22}\right)
\end{equation}
where $\sigma_{L,22}$ and $\sigma_{R,22}$ are defined, for our model, in
eq.~\rf{eq:RHL} and eq.~\rf{eq:RS}. The sum in eq.~\rf{eq:RHL} should
run only over the heavy neutral leptons, as the light neutrino
contribution is alreday included in the Standard Model. We have
checked that our formulas reproduced the PDG~\cite{Amsler:2008zzb}
value for the sum of the $W$ and $Z$ diagrams, at 1-loop order in the
Standard Model.

%-----------------------------------------------------------------------------

%----------------------------------------------------------------------
\end{document}